\documentclass[12pt,onecolumn]{article}
\usepackage[margin=2.5cm]{geometry}

\usepackage{amsmath,amssymb,amsthm} 
\usepackage{color}
\usepackage{enumitem}
\usepackage{authblk}
\usepackage{graphicx}
\usepackage{hyperref}
\usepackage{empheq}

\newcommand{\AK}{\textcolor{black}}
\newcommand{\AKn}{\textcolor{black}}

\begin{document}
\title{Focusing of Active Particles in a Converging Flow}

\author[1]{Mykhailo Potomkin}
\author[2]{Andreas Kaiser}
\author[1]{Leonid Berlyand}
\author[1,3,*]{Igor Aranson}

\affil[1]{Department of Mathematics, Pennsylvania State University, University Park, 16803, USA}
\affil[2]{Materials Science Division, Argonne National Laboratory, 9700 South Cass Avenue, Argonne, Illinois 60439, USA}
\affil[3]{Department of Biomedical Engineering, Pennsylvania State University, University Park, 16803, USA}

\affil[*]{corresponding author: isa12@psu.edu}



\maketitle


\begin{abstract}
We consider active particles swimming in a convergent fluid flow in a trapezoid nozzle with no-slip walls. 
We use mathematical modeling to analyze trajectories of these particles inside the nozzle.
By extensive Monte Carlo simulations, we show that trajectories are strongly affected by the background fluid flow and geometry of the nozzle leading to wall accumulation and upstream motion (rheotaxis). In particular, we describe the non-trivial focusing
of active rods depending on physical and geometrical parameters. It is also established that the convergent component of the background flow leads to stability of both downstream and upstream swimming at the centerline. The stability of downstream swimming enhances focusing, and the stability of upstream swimming enables rheotaxis in the bulk.
\end{abstract}


\section{Introduction}

Active matter consists of a large number of self-driven agents converting chemical energy, usually stored in the surrounding environment, into mechanical motion \cite{Ram2010,MarJoaRamLivProRaoSim2013,ElgWinGom2015}. 
In the last decade various realizations of active matter have been studied including living self-propelled particles as well as synthetically manufactured ones. Living agents are for example bacteria \cite{DomCisChaGolKes2004,sokolov2007concentration}, microtubules in biological cells \cite{SurNedLeiKar2001,SanCheDeCHeyDog2012}, spermatozoa \cite{2005Riedel_Science,Woolley,2008Friedrich_NJP} and animals \cite{CavComGiaParSanSteVia2010,CouKraJamRuxFra2002,VisZaf2012}. 

Such systems are out-of-equilibrium and show a variety of collective effects, from clustering \cite{BocquetPRL12,Bialke_PRL2013,Baskaran_PRL2013,Palacci_science} 
to swarming, swirling and turbulent type motions \cite{ElgWinGom2015, DomCisChaGolKes2004,sokolov2007concentration,VisZaf2012,wensink2012meso,SokAra2012,SaiShe08,RyaSokBerAra13}, reduction of effective viscosity \cite{sokolov2009reduction,GacMinBerLinRouCle2013,LopGacDouAurCle2015,HaiAraBerKar08,HaiSokAraBer09,HaiAroBerKar10,RyaHaiBerZieAra11}, extraction of useful energy \cite{sokolov2010swimming,di2010bacterial,kaiser2014transport}, 
and enhanced mixing \cite{WuLib2000,SokGolFelAra2009,pushkin2014stirring}. 
Besides the behavior of microswimmers in the bulk the influence of confinement has been studied intensively in experiments \cite{DenissenkoPNAS,Chaikin2007} and numerical simulations \cite{ElgetiGompper13,Lee13Wall,Ghosh,Wensink2008}.
There are two distinguishing features of swimmers confined by walls and exposed to an external flow: accumulation at the walls and upstream motion (rheotaxis). Microorganisms such as bacteria \cite{BerTur1990,RamTulPha1993,FryForBerCum1995,VigForWagTam2002,BerTurBerLau2008} and  sperm cells \cite{Rot1963} are typically attracted by no-slip surfaces. Such accumulation was also observed for larger organisms such as worms \cite{YuaRaiBau2015} and for synthetic particles \cite{DasGarCam2015}. The propensity of active particles to turn themselves against the flow (rheotaxis) is also typically observed. \AKn{While for larger organisms, such as fish, rheotaxis is caused by a deliberate response to a stream to hold their position 
\cite{JiaTorPeiBol2015}, for micron sized swimmers rheotaxis has a pure mechanical origin \cite{HilKalMcMKos2007,fu2012bacterial,YuaRaiBau2015rheo,TouKirBerAra14,PalSacAbrBarHanGroPinCha2015}.}

These phenomena observed in living active matter can also be achieved using synthetic swimmers, such as self-thermophoretic \cite{Sano_PRL2010} and self-diffusiophoretic \cite{paxton2004catalytic,HowsePRL2007,Bechinger_JPCM,Baraban_SM2012} micron sized particles as well as particles set into active motion due to the influence of an external field \cite{bricard2013emergence,bricard2015emergent,KaiserSciAdv2017}.

Using simple models we describe the extrusion of a dilute active suspension through a trapezoid nozzle. 
We analyze the qualitative behavior of trajectories of an individual active particle in the nozzle and study the statistical properties of the particles in the nozzle. 
The accumulation at walls and rheotaxis are important for understanding how an active suspension is extruded through a nozzle. Wall accumulation may eliminate all possible benefits caused by the activity of the particles in the bulk. 
  \AKn{Due to rheotaxis active particles may never reach the outlet and leave the nozzle through the inlet, so that properties of the suspension coming out through the outlet will not differ from those of the background fluid.}

The specific geometry of the nozzle is also important for our study. The nozzle is a finite domain with two open ends (the inlet and the outlet) and the walls of the nozzle are not parallel but convergent, that is, the distance between walls decreases from the inlet to the outlet. The statistical properties of active suspension (e.g., concentration of active particles) extruded in the infinite channel with parallel straight or periodic walls are well-established, see e.g., \cite{EzhSai2015} and \cite{MalSta2017}, respectively. The finite nozzle size leads to a ``proximity effect", i.e., the equilibrium distribution of active particles changes significantly in proximity of both the inlet and the outlet. The fact that the walls are convergent, results in a ``focusing effect", i.e., the background flow compared to the pressure driven flow in the straight channel (the Poiseuille flow) has an additional convergent component that turns a particle toward the centerline.  \AK{Specifically, in this work it is shown that due to this convergent component of the background flow both up- and downstream swimming at the centerline are stable. Stability of the upstream swimming at the centerline is somewhat surprising since from observations in the Poisueille flow it is expected that an active particle turns against the flow only while swimming towards the walls, where the shear rate is higher. This means that we find rheotaxis in the bulk of an active suspension.}


\section{Model}
\label{sec:model}

\begin{figure}[ht]
	\centering
	\includegraphics[width=0.5\textwidth]{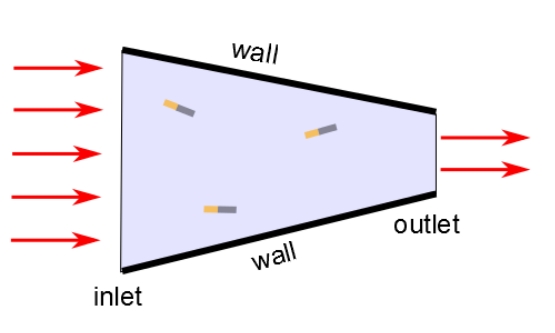}
	\caption{Sketch of a trapezoid nozzle filled with an dilute suspension of rodlike active particles in the presence of a converging background flow.}
	\label{fig:nozzle-sketch}
\end{figure}

To study the dynamics of active particles in a converging flow, two modeling approaches are exploited. In both, an active particle is represented by a rigid rod of length $\ell$ swimming in the $xy$-plane. In the first - simpler - approach, the rod is a one-dimensional segment which cannot penetrate a wall, whereas in the second - more sophisticated - approach we use the Yukawa segment model \cite{Kirchhoff1996} to take into account both finite length and width of the rod, as well as a more accurate description of particle-wall steric interaction.

The active particle's center location and its unit orientation vector are denoted by ${\bf r}=(x,y)$ and ${\bf p}=(\cos \varphi,\sin\varphi)$, respectively. The active particles are self-propelled with a velocity directed along their orientation $v_{0}{\bf p}$. 
\AK{The active particles are confined by a nozzle, see Fig.~\ref{fig:nozzle-sketch}, which is an isosceles trapezoid $\Omega$, placed in the $xy$-plane so that inlet $x=x_{\text{in}}$ and outlet  $x=x_{\text{out}}$ are bases and the $y$-axis is the line of symmetry:
\begin{equation}
\Omega=\left\{x_{\text{in}}<x<x_{\text{out}}, \; \alpha^2 x^2 -y^2>0\right\}.
\end{equation}
The nozzle length, the distance between the inlet and the outlet, is denoted by $L$, i.e., $L=|x_{\text{out}}-x_{\text{in}}|$. The width of the outlet and the inlet are denoted by $w_{\text{out}}$ and $w_{\text{in}}$, respectively, and their ratio is denoted by $k={w_{\text{out}}}/{w_{\text{in}}}$.}

\AK{Furthermore, the active particles are exposed to an external background flow. We approximate the resulting converging background flow due to the trapezoid geometry of the nozzle by
\begin{equation}\label{convergent_flow}
{\bf u}_{\text{BG}}({\bf r})=(u_x(x,y),u_y(x,y))=(-u_0 (\alpha^2 x^2-y^2)/x^3, -u_0 y (\alpha^2x^2-y^2)/x^4),
\end{equation}
where $u_0$ is a constant coefficient related to the flow rate and $\alpha$ is the slope of walls of the nozzle. 
Equation \eqref{convergent_flow} is an extension of the Poiseuille flow to channels with convergent walls\footnote{In order to recover the Poiseuille flow (for channels of width $2H$) from Eq.~\eqref{convergent_flow}, take $x=H/\alpha$, $u_0=H^3/\alpha^3$ and pass to the limit $\alpha\to 0$. Note that the walls of the nozzle are placed so that they intersect at the origin, so in the limit of parallel walls, $\alpha \to 0$, both the inlet and the outlet locations, $x_{\text{in}}$ and $x_{\text{out}}$, go to $-\infty$.}}.


Active particles swim in the low Reynolds-number regime. The corresponding overdamped equations of motion for the locations  ${\bf r}$ and orientations ${\bf p}$ are given by:
\begin{equation}
\label{orig-location}
\dfrac{\text{d}\bf r}{\text{d}t}={\bf u}_{\text{BG}}({\bf r})+v_{0}{\bf p},
\end{equation}
\begin{equation}
\label{orig-orientation}
\dfrac{\text{d}\bf p}{\text{d}t} =(\text{I}-{\bf p}{\bf p}^{\text{T}})\nabla_{\bf r}{\bf u}_{\text{BG}}({\bf r}){\bf p}\,+\sqrt{2D_r}\,\zeta \,{\bf e}_{\varphi}.
\end{equation}
Here \eqref{orig-orientation} is the Jeffery's equation \cite{SaiShe08,Jef1922,KimKar13} for rods with an additional term due to random re-orientation with rotational diffusion  coefficient $D_r$; $\zeta$ is an uncorrelated noise with the intensity $\langle \zeta(t),\zeta(t')\rangle=\delta(t-t')$, $e_{\varphi}=(-\sin \varphi, \cos \varphi)$.  Equation \eqref{orig-orientation} can also be rewritten for the orientation angle $\varphi$:
\begin{equation}
\label{orig-orientation-angle}
\dfrac{\text{d}\varphi}{\text{d}t}=\omega+ \nu\, \sin 2\varphi + \gamma \,\cos 2\varphi+\sqrt{2D_r}\,\zeta.
\end{equation}
Here $\omega=\dfrac{1}{2}\left(\dfrac{\partial u_y}{\partial x}-\dfrac{\partial u_x}{\partial y}\right)$, $\nu=\dfrac{1}{2}\left(\dfrac{\partial u_y}{\partial y}-\dfrac{\partial u_x}{\partial x}\right)=\dfrac{\partial u_y}{\partial y}=-\dfrac{\partial u_x}{\partial x}$, and $\gamma=\dfrac{1}{2}\left(\dfrac{\partial u_y}{\partial x}+\dfrac{\partial u_x}{\partial y}\right)$ are local vorticity, vertical expansion (or, equivalently, horizontal compression; similar to Poisson's effect in elasticity) and shear.  

The strength of the background flow is
\AK{quantified by the inverse Stokes number, which is the ratio between the background flow at the center of the inlet and the self-propulsion velocity $v_0$. Specifically,}
\begin{equation}
\sigma = \dfrac{u_x(x_{\text{in}},0)}{v_{0}}=\dfrac{u_0\alpha^2}{v_{0}|x_{\text{in}}|},
\end{equation}    
where $(x_{\text{in}},0)$ denotes the location at the center of the inlet.



In the first modeling approach we include the particle wall interaction in the following way: an active particle is not allowed to penetrate the walls of the nozzle. To enforce this, we require that both the front and the back of the particle, ${\bf r}(t)\pm(\ell/2) {\bf p}$, are located inside the nozzle. In numerical simulations of the system \eqref{orig-location}-\eqref{orig-orientation-angle} this requirement translates into the following rule: if during numerical integration of \eqref{orig-location}-\eqref{orig-orientation-angle} a particle penetrates one of the two walls, then this particle is instantaneously shifted back along the inward normal at the minimal distance, so its front and back are again located inside the nozzle while its orientation is kept fixed.

\AKn{Unless mentioned otherwise, in this modeling approach we consider a nozzle whose inlet width $w_{\text{in}}=0.2$ mm and outlet width $w_{\text{out}}=0.1$ mm are fixed. The following nozzle lengths are considered:  $L=0.2$ mm, $L=0.5$ mm and $L=1.0$ mm. The length of the active particles is $\ell = 20$ $\mu$m, they swim with a self-propulsion velocity $v_{0}=10$ $\mu$m $\text{s}^{-1}$ and their rotational diffusion coefficient is given by $D_r=0.1$ $\text{s}^{-1}$.}

All active particles are initially placed at the inlet, $x(0)=x_{\text{in}}$, with random $y$-component $y(0)$ and orientation angle $\varphi(0)$. The probability distribution function for initial conditions $y(0)$ and $\varphi(0)$ is given by $\Psi\propto 1$ (uniform). The trajectory of an active particle is studied until it leaves the nozzle either through the inlet or the outlet. To gather statistics we use 96,000 trajectories.


\begin{figure}[ht!]
	\centering
	\includegraphics[width=0.6\textwidth]{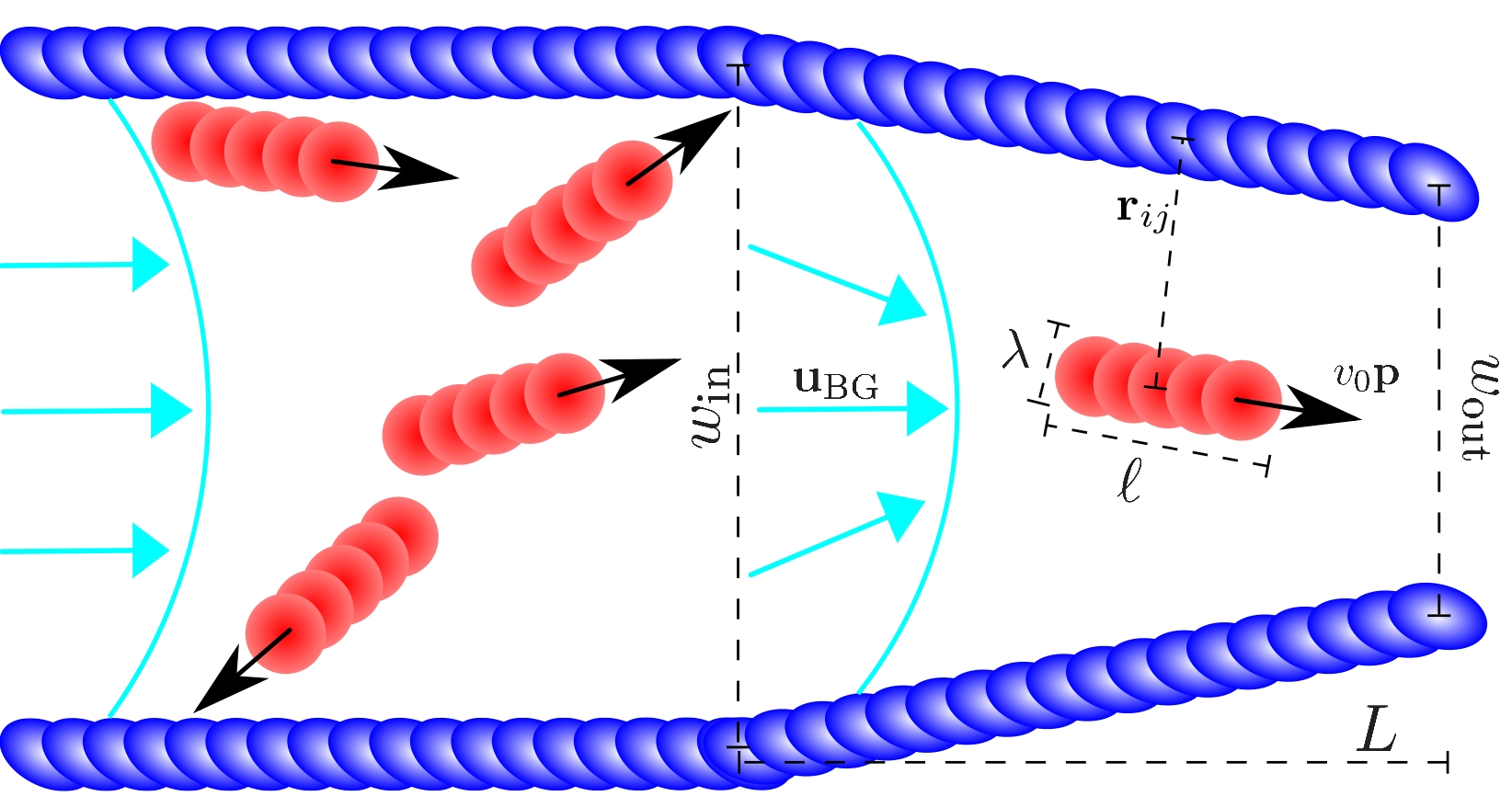}
	\caption{Sketch of a discretized active rod (red) of length $\ell$ and width $\lambda$ which is propelled with a velocity $v_0$ along its orientation ${\bf p}$ and is exposed to a converging background flow ${\bf u}_{\text{BG}}$ in the presence of a trapezoid nozzle confinement of length $L$ and with an inlet of size $w_{\text{in}}$ and outlet of size $w_{\text{out}}$(blue).
	To study a system with a packing fraction $\rho=0.1$ a channel is attached to the inlet with a non-converging background flow.}
	\label{fig:nozzle-sketchAK}
\end{figure}

 We use the second approach to describe the particle-wall interactions and the torque induced by the flow more accurately. \AKn{For this purpose each rod, representing an active particle, of length $\ell$, width $\lambda$ and the corresponding aspect ratio $a=\ell/\lambda$ is discretized into $n_r$ spherical segments with $n_r = \lfloor 9 a /8 \rceil$ ($\lfloor x \rceil$ denotes the nearest integer function).} The resulting segment distance is also used to discretize the walls of the nozzle into $n_w$ segments in the same way. Between the segments of different objects a repulsive Yukawa potential is imposed. The resulting total pair potential is given by $U = U_0\sum_{i=1}^{n_r}\sum_{j=1}^{n_w} \exp [-r_{ij} / \lambda]/r_{ij}$, where $\lambda$ is the screening length defining the particle diameter, $U_0$ is the prefactor of the Yukawa potential and $r_{ij} = |{\bf r}_{i} - {\bf r}_{j}|$ is the distance between segment $i$ of a rod and $j$ of the wall of the nozzle, see Fig.~\ref{fig:nozzle-sketchAK}.

The equations of motion (\ref{orig-location}) and (\ref{orig-orientation}) are complemented by the respective derivative of the total potential energy of a rod along with the one-body translational and rotational friction tensors for the rods ${\bf f}_{\cal T}$ and ${\bf f}_{\cal R}$ which can be decomposed into parallel $f_\parallel$, perpendicular $f_\perp$
and rotational $f_{\cal R}$ contributions which depend solely on the aspect ratio $a$~\cite{tirado}.
For this approach we measure distances in units of $\lambda$, velocities in units of $v_0=F_0/f_\parallel$ (here $F_0$ is an effective self-propulsion force), and time in units of $\tau = \lambda f_\parallel / F_0$. While the width of the outlet $w_{\text{out}}$ is varied, the width of the inlet $w_{\text{in}}$ as well as the length of the nozzle $L$ is fixed to $100\lambda$ in our second approach. 
\AKn{Initial conditions are the same as in the first approach. 
To avoid that a rod and a wall initially intersect each other, the rod is allowed to reorient itself during an equilibration time $t_e = 10 \tau$  while its center of mass is fixed.}

\AKn{Furthermore, we use the second approach to study the impact of a finite density of swimmers. For this approach we initialize $N$ active rods in a channel confinement which is connected to the inlet of the nozzle, see Fig.~\ref{fig:nozzle-sketchAK}. Inside the channel we assume a regular (non-converging) Poiseuille flow~\cite{zottl2013periodic}. We restrict our study to a dilute active suspension with a two dimensional packing fraction $\rho=0.1$. To maintain this fraction, particles which leave the simulation domain are randomly placed at the inlet of the channel confinement.}


\section{Results}
\label{sec:results}


\subsection{Focusing of outlet distribution}
\label{sec:focusing}

Here we characterize the properties of the particles leaving the nozzle at either the outlet or the inlet. Specifically, our objective is to determine whether particles accumulate at the center or at walls when they pass through the outlet or the inlet.

\begin{figure}[h!]
	\centering
	\includegraphics[width=1.0\textwidth]{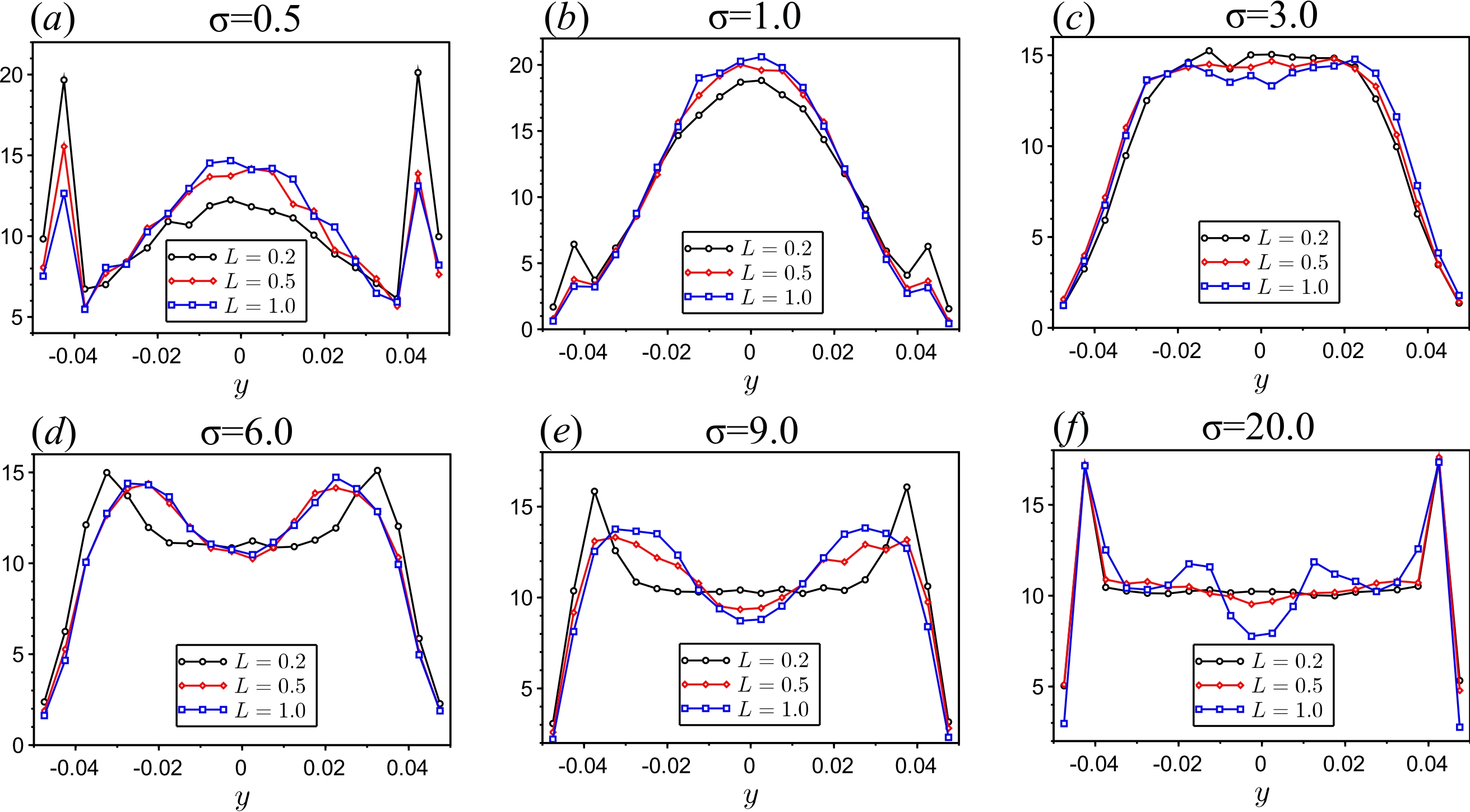}
	\caption{Histograms of the outlet distribution for $y|_{\text{out}}$ for given  \AK{inverse Stokes number} $\sigma$ and length $L$ of the nozzle. The histograms are obtained from numerical integration of \eqref{orig-location}-\eqref{orig-orientation-angle}. 
		 }
	\label{fig:dependence-on-sigma}
\end{figure}

\begin{figure}[h!]
	\centering
	\includegraphics[width=1.0\textwidth]{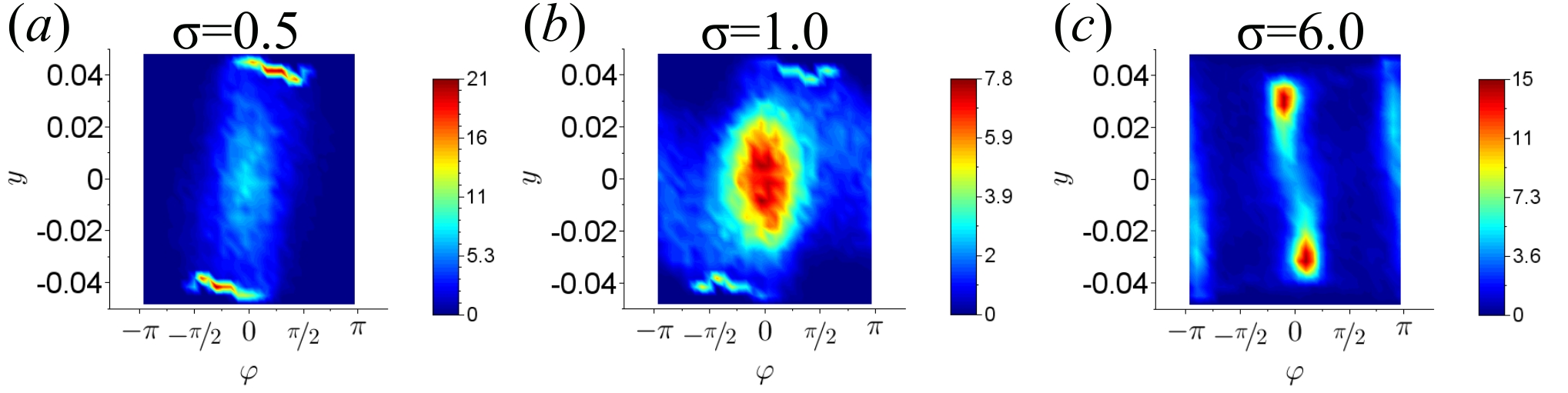}
	\caption{Outlet distribution histograms for $(y,\varphi)|_{\text{out}}$ computed for given inverse Stokes number $\sigma$ and nozzle length $L=0.2$. }
	\label{fig:y_phi_diagram}
\end{figure}

We start with the first modeling approach. Figure~\ref{fig:dependence-on-sigma} shows the spatial distribution of active particles leaving the nozzle at the outlet for various \AK{inverse Stokes number} $\sigma$ and three different lengths $L$ of the nozzle, while the width of the inlet and the outlet are fixed. 
For small inverse Stokes number $\sigma$, the background flow is negligible compared to the self-propulsion velocity. Active particles swim close to the walls and peaks at walls are still clearly visible for $\sigma=0.5$ for all nozzle lengths $L$, see Fig.~\ref{fig:dependence-on-sigma}(a). For $\sigma=1$, the self-propulsion velocity and the background flow are comparable; in this case the histogram shows a single peak at the center of the outlet, see Fig.~\ref{fig:dependence-on-sigma}(b). Further increasing the inverse Stokes number from $\sigma=1$ to $\sigma=9$ leads to a broadening of the central peak and then to the formation of two peaks with a well in the center of the outlet, see Fig.~\ref{fig:dependence-on-sigma}(c)-(e). Finally, for an even larger inverse Stokes number $\sigma$,  the self-propulsion velocity is negligible and the histogram becomes close to the one in the passive (no self-propulsion, $v_0 = 0$) case, see Fig.~\ref{fig:dependence-on-sigma}(f). Here the histogram for a nozzle length $L=0.2$ mm is uniform except at the edges where it has local peaks due to accumulation at the walls caused by steric interactions.

Histograms for both the $y$-component and the orientation angle $\varphi$ of the active particles reaching the outlet are depicted in Fig.~\ref{fig:y_phi_diagram}(a)-(c). While active particles leave the nozzle with orientations away from the centerline for small inverse Stokes number, $\sigma = 0.5$, they are mostly oriented towards the centerline for larger values of the inverse Stokes number.
\AK{In Fig.~\ref{fig:y_phi_diagram}(c), one can observe that the histogram is concentrated largely for downstream orientations $\varphi \approx 0$ and slightly for upstream orientations $\varphi \approx \pm \pi$. These local peaks for $\varphi \approx \pm \pi$ away from walls are evidence of rheotaxis in the bulk. \AKn{These peaks are visible for large inverse Stokes numbers only and the corresponding active particles are flushed out of the nozzle with upstream orientations.}
	}

\begin{figure}[ht!]
	\centering
	\includegraphics[width=1.0\textwidth]{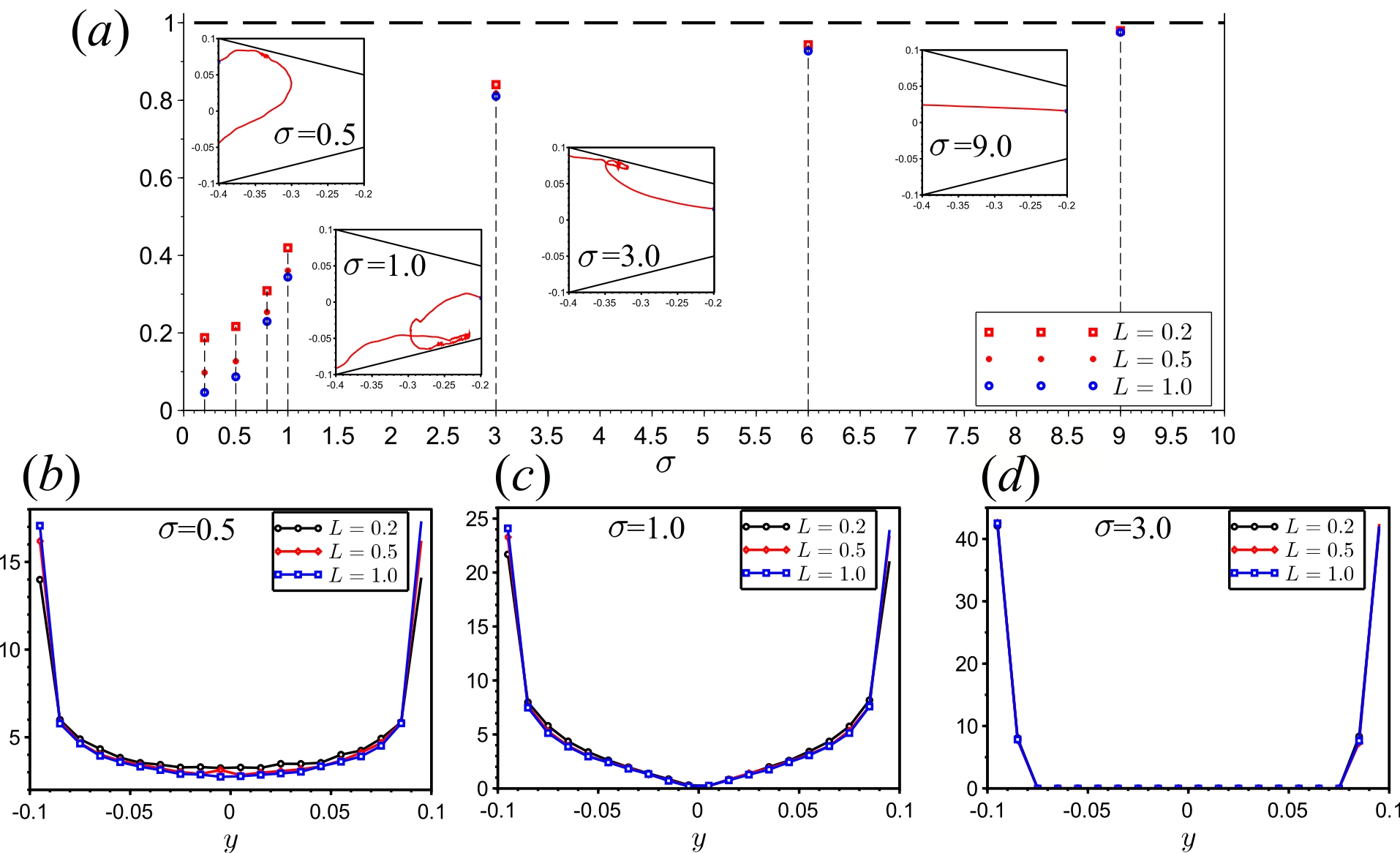}
	\caption{(a) Probability of active particles to reach the outlet for various \AK{inverse Stokes number} $\sigma$ (horizontal axis) and given lengths of the nozzle $L$. Insets: Trajectories for the case of $L=0.2$ mm.  (b-d) Distribution histograms for particles leaving the nozzle through the inlet $y|_{\text{in}}$ computed for given reduced flow velocities $\sigma$ and nozzle lengths $L$.
	}
	\label{fig:share-of-particles}
\end{figure}

\AKn{Due to rotational diffusion and rheotaxis it is possible that an active particle can leave the nozzle through the inlet. We compute the probability of active particles to reach the outlet.} This probability, as a function of the inverse Stokes number $\sigma$ for the three considered nozzle lengths $L$, is shown in Fig.~\ref{fig:share-of-particles}(a), together with selected trajectories, see insets in Fig.~\ref{fig:share-of-particles}(a). 
The figure shows that the probability that an active particle eventually reaches the outlet monotonically grows with \AK{the inverse Stokes number} $\sigma$. Note that a passive  particle always leaves the nozzle through the outlet. By comparing the probabilities for different nozzle lengths $L$ it becomes obvious that an active particle is less likely to leave the nozzle through the outlet for longer nozzles. Due to the larger distance $L$ between the inlet and the outlet an active particle spends more time within the nozzle, which makes it more likely to swim upstream by either rotational diffusion or rheotaxis. 
In Fig.~\ref{fig:share-of-particles}(b)-(d) histograms for active particles leaving the nozzle through the inlet are shown. In the case of small inverse Stokes number, $\sigma=0.5$, the majority of active particles leaves the nozzle at the inlet. Specifically, most of them swim upstream due to rheotaxis close to the walls, but some active particles leave the nozzle at the inlet close to the center. These active particles are oriented upstream due to random reorientation. By increasing the inverse Stokes number $\sigma \geq 1$, active particles  are no longer able to leave the nozzle at the inlet close to the center.

\begin{figure}[ht]
	\centering
	\includegraphics[width=0.5\textwidth]{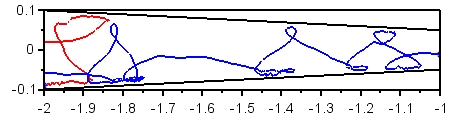}
	\caption{Examples of two trajectories for $L=1$ mm and $\sigma=1.0$. The red trajectory starts and ends at the inlet (the endpoint is near the lower wall). The blue trajectory has a zigzag shape with loops close to the walls; the particle that corresponds to the blue trajectory manages to reach the outlet.}
	\label{fig:two_representative}
\end{figure}

Let us now consider specific examples of active particles' trajectories, 
see
Fig.~\ref{fig:two_representative}. The first trajectory (red) starts and ends at the inlet. 
Initially the active particle swims downstream and collides with the upper wall due to the torque induced by the background flow. Close to the wall it exhibits rheotactic behavior, but before it reaches the inlet it is expelled towards the center of the nozzle due to rotational diffusion, similar to bacteria that may escape from surfaces due to tumbling \cite{DreDunCisGanGol2011}. 
Eventually, the active particle leaves the nozzle at the inlet.
As for the other depicted trajectory (blue), the active particle manages to reach the outlet. Along its course through the nozzle it swims upstream several times but in the end the active particle is washed out through the outlet by the background flow. For larger flow rates the trajectories of active particles are less curly, since the flow gets more dominant, see insets of Fig.~\ref{fig:share-of-particles}(a).



\begin{figure}[ht]
	\centering
	\includegraphics[width=1.0\textwidth]{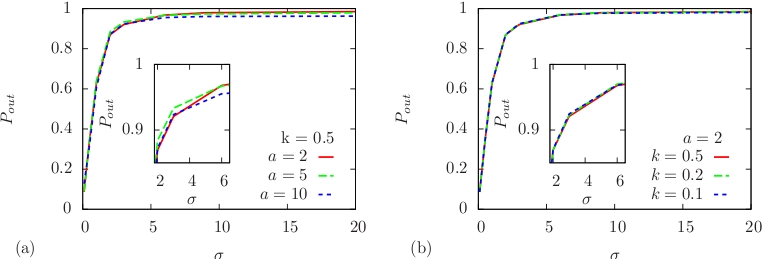}
	\caption{(a) Probability for an active particle to reach the outlet of the nozzle $P_{\text{out}}$ as a function of \AK{inverse Stokes number} $\sigma$ for three given aspect ratios $a$ of self-propelled rods and (b) for a fixed aspect ratio $a$ and three given ratios of the nozzle $k$. \AK{Insets show close-ups.}}
	\label{fig:AKratio}
\end{figure}

\begin{figure}[t!]
	\centering
	\includegraphics[width=1.0\textwidth]{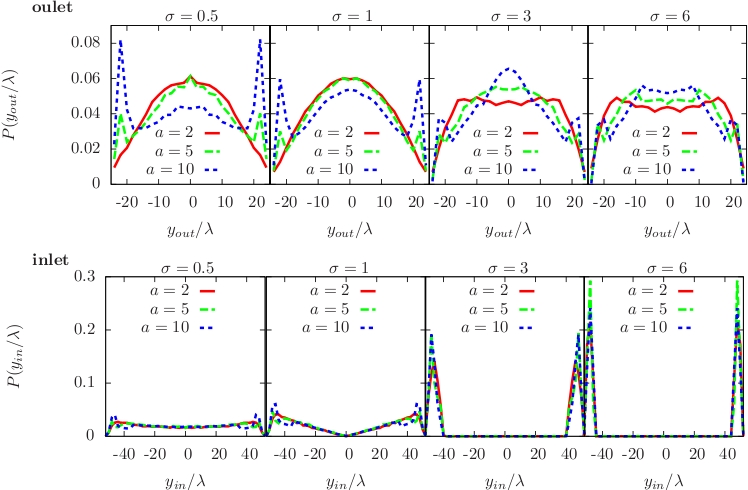}
	\caption{ Comparison of the spatial distribution of active particles at (top row) the outlet and (bottom row) the inlet of the nozzle for given inverse Stokes numbers $\sigma$ and aspect ratios $a$, an outlet width $w_{\text{out}}=50\lambda$ and an inlet width $w_{\text{in}}=100\lambda$.}
	\label{fig:AK1d}
\end{figure}

Next we present results of the second modeling approach which is based on the Yukawa-segment model.
So far we have concentrated on fixed widths of the inlet and outlet. Here we consider nozzles with fixed length $L$ and inlet width $w_{\text{in}}$ and vary 
nozzle ratio $k$. We study the behavior of active rods with varied aspect ratio $a$. 
As shown in Fig.~\ref{fig:AKratio}, neither the aspect ratio $a$, see Fig.~\ref{fig:AKratio}(a),  nor the nozzle ratio $k$,  see Fig.~\ref{fig:AKratio}(b),  have a significant impact on the probability $P_{\text{out}}$ which measures how many active rods leave the nozzle at the outlet. However, the aspect ratio $a$ is important for the location where the active rods leave the nozzle at the inlet and the outlet, see Fig.~\ref{fig:AK1d}. For short rods $(a=2)$ and small inverse Stokes numbers $(\sigma \leq 1)$ the distribution of active particles shows just a single peak located at the center. This peak broadens if the inverse Stokes number increases, which is in perfect agreement with the results obtained by the first approach, cf. Fig.~\ref{fig:dependence-on-sigma}. 
It is more likely for short rods than for long ones to be expelled towards the center due to rotational diffusion.
Hence the distribution of particles at the outlet for long rods $(a=10)$ shows additional peaks close to the wall. These peaks become smaller if the inverse Stokes number increases. The distribution of particles leaving the nozzle at the inlet is similar to our first approach. While the distribution is almost flat for small inverse Stokes numbers, increasing this number makes it impossible to leave the nozzle close to the center at the inlet. Similar to the outlet the wall accumulation at the inlet is more pronounced for longer rods.

\begin{figure}[h!]
	\centering
	\includegraphics[width=1.0\textwidth]{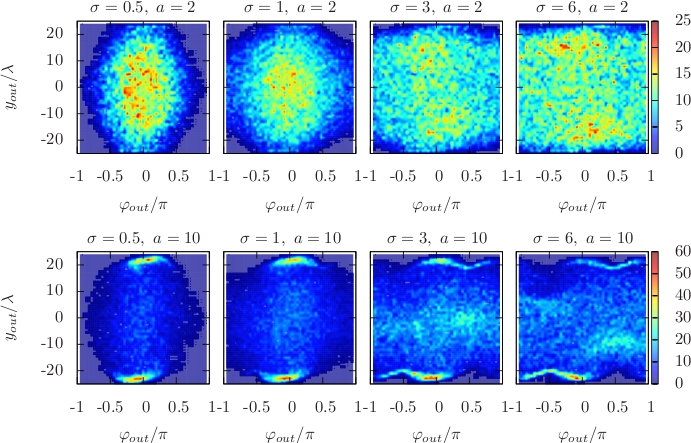}
	\caption{Outlet distribution histograms for $(y,\varphi)|_{\text{out}}$ computed for given \AK{inverse Stokes numbers} $\sigma$ and a nozzle with an outlet width of $w_{\text{out}}=50\lambda$ for active rods with an aspect ratio (top row) $a=2$  and (bottom row) $a=10$.}
	\label{fig:AK2d}
\end{figure}

By comparing the orientation of the particles at the outlet, the influence of the actual length of the rod becomes visible, see Fig.~\ref{fig:AK2d}. As seen before for short rods, $a=2$, for small inverse Stokes numbers $\sigma$ there is no wall accumulation. Hence most particles leave the nozzle close to the center and are orientated in the direction of the outlet. This profile smears out if the inverse Stokes number is increased to $\sigma = 1$. For  larger inverse Stokes numbers the figures are qualitatively similar to the one obtained by the first approach, cf. Fig.~\ref{fig:y_phi_diagram}(c). Particles in the bottom half of the nozzle tend to point upwards and particles in the top half tend to point downwards.
The same tendency is seen for long rods $a=10$ and small inverse Stokes number. However for long active rods, this is because they slide along the walls. The bright spots close to the walls for long rods and large inverse Stokes numbers indicate that particles close to the walls 
are flushed through the outlet by the large background flow even if they are oriented upstream.
\AK{In addition, there are blurred peaks away from the walls for large inverse Stokes numbers $\sigma$. The corresponding particles crossed the outlet with mostly upstream orientations. This is similar to Fig.~\ref{fig:y_phi_diagram}(c), where particles exhibiting in-bulk rheotactic characteristics were observed at the outlet of the nozzle.}

\begin{figure}[h!]
	\centering
	\includegraphics[width=1.0\textwidth]{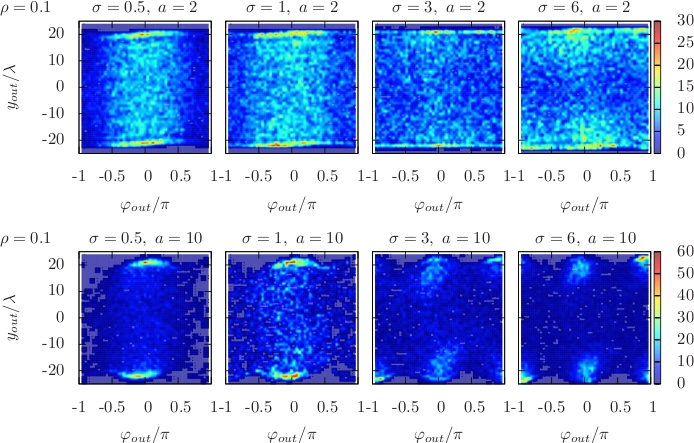}
	\caption{Outlet distribution histograms for $(y,\varphi)|_{\text{out}}$ computed for given \AK{inverse Stokes numbers} $\sigma$ and a nozzle with an outlet width  $w_{\text{out}}=50\lambda$ for active rods with an aspect ratio (top row) $a=2$  and (bottom row) $a=10$ for a packing fraction $\rho = 0.1$. 
	}
	\label{fig:AK2dint}
\end{figure}

\AK{By comparing the results for individual active rods, see again Fig.~\ref{fig:AK2d}, with those for interacting active rods at a finite packing fraction $\rho = 0.1$, see Fig.~\ref{fig:AK2dint}, we find that wall accumulation becomes more pronounced. Mutual collisions of the rods lead to a broader distribution of particles. For long rods, $a=10$, the peaks at $\varphi \approx 0$ and $\varphi \approx \pm \pi$ remain close to the walls and the blurred peaks at the center vanish. }  
   
\subsection{Optimization of focusing}
\label{sec:optimization}

Here we study the properties of the active particles in more detail and provide insight into the nozzle geometry, the background flow and the size of the swimmers that should be used in order to optimize the focusing at the outlet of the nozzle. 


For this purpose we study three distinct quantities. The averaged dwell time $\langle T\rangle$, the time it takes for an active particle to reach the outlet, the mean alignment of the particles measured by $\langle \cos \varphi_{\text{out}}\rangle$ and the mean deviation from the center $y=0$ at the outlet $\langle |y_{\text{out}}|\rangle$.
As depicted in Fig.~\ref{fig:share-of-particles}, for increasing inverse Stokes number the probability for active particles to reach the outlet increases. However they are spread all over the outlet. This is quantified by the $\langle |y_{\text{out}}|\rangle$. Small values of $\langle |y_{\text{out}}|\rangle$ correspond to a better focusing. If particles leave the nozzle with no preferred orientation, their mean orientation vanishes, $\langle \cos \varphi_{\text{out}}\rangle = 0$; in case of being orientated upstream we obtain $\langle \cos \varphi_{\text{out}}\rangle = -1$ and finally $\langle \cos \varphi_{\text{out}}\rangle = 1$ if the particles are pointing in the direction of the outlet. Obviously in an experimental realization a fast focusing process and hence small dwell times $T$ would be preferable. 

\begin{figure}[ht]
	\centering
	\includegraphics[width=1.0\textwidth]{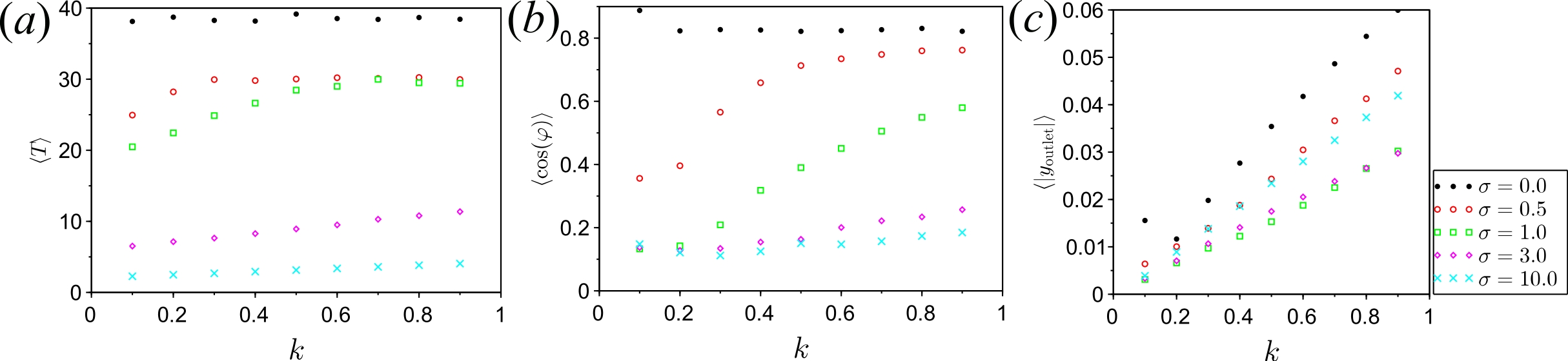}
	\caption{(a) Dwell time $\langle T\rangle$; (b) mean alignment at the outlet, $\langle \cos \varphi\rangle$; (c) mean deviation from center $y=0$ at the outlet $\langle |y_{\text{out}}|\rangle$. 
		}
	\label{fig:optimization}
\end{figure}

The numerical results obtained by the first modeling approach are depicted in Fig.~\ref{fig:optimization}. While the dwell time hardly depends on the size ratio $k$ of the nozzle, obviously the strength of the background flow has a huge impact on the dwell time and large inverse Stokes numbers $\sigma$ lead to a faster passing through the nozzle of the active particles, see Fig.~\ref{fig:optimization}(a). The alignment of the active particles, $\langle \cos \varphi_{\text{out}}\rangle$, becomes better if the nozzle ratio $k$ is large and the flow is slow, see Fig.~\ref{fig:optimization}(b). The averaged deviation from the centerline $\langle |y_{\text{out}}|\rangle$ increases with increasing nozzle ratio $k$ since the width of the outlet becomes larger. As could already be seen in Fig.~\ref{fig:dependence-on-sigma}, the averaged deviation from the centerline is non-monotonic as a function of the inverse Stokes number and shows the smallest distance from the centerline for all nozzle ratios if the strength of the flow is comparable to the self-propulsion velocity of the swimmers, $\sigma=1$.


\begin{figure}[h]
	\centering
	\includegraphics[width=1.0\textwidth]{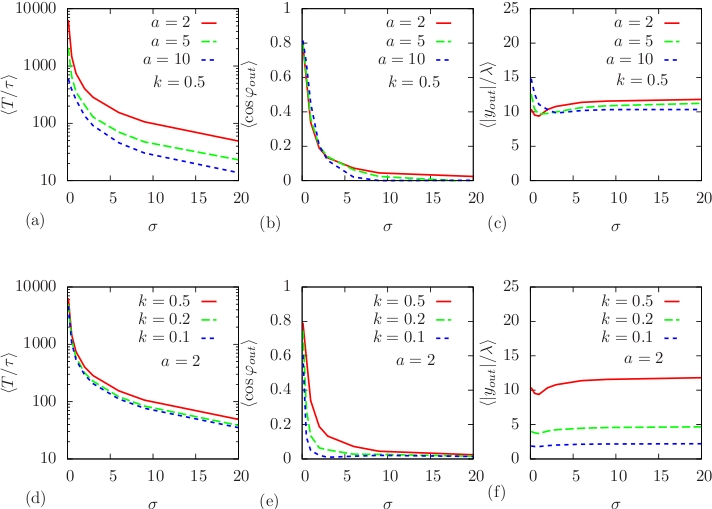}
	\caption{(a,e) Dwell time $\langle T\rangle$, (b,f) the mean alignment, $\langle \cos \varphi_{\text{out}}\rangle$ and (c,f) mean deviation from center $y=0$ at the outlet $\langle |y_{\text{out}}|\rangle$ for (top row) a fixed outlet width of $w_{\text{out}}=50\lambda$ and given aspect ratios $a$ of the swimmers and (bottom row) fixed aspect ratio $a=2$ and varied nozzle ratio $k$, whereby the width of the outlet changes.}
	\label{fig:AKopti}
\end{figure}

Let us now study how these three quantities depend on the aspect ratio of the swimmer. To this end, we use the second modeling approach. We consider all three parameters as a function of the \AK{inverse Stokes number} $\sigma$. Longer rods have a shorter dwell time so that they reach the outlet faster, see Fig.~\ref{fig:AKopti}(a). Increasing the flow velocity obviously leads to a decreasing dwell time. The same holds for the mean alignment --  it decreases for increasing inverse Stokes number, see Fig.~\ref{fig:AKopti}(b). Moreover, for small inverse Stokes numbers, $\sigma \leq 2$, the mean alignment is better for long rods. For large inverse Stokes numbers,  long rods $a=10$ are washed out with almost random orientation, however short rods $a=2$ are slightly aligned with the flow. Short rods are focused better for small inverse Stokes numbers, $\sigma \leq 2$, see Fig.~\ref{fig:AKopti}(c),  due to wall alignment and wall accumulation of longer rods. For larger inverse Stokes numbers, it is the other way around -- long rods are better focused.
Comparing various nozzle ratios $k$ with fixed simmers' aspect ratio $a$, we obtain that smaller ratios $k$ lead to smaller dwell times [Fig.~\ref{fig:AKopti}(d)] and better alignment [Fig.~\ref{fig:AKopti}(e)]. For narrow outlets (small $k$) the active particles leave the outlet closer to the center, see Fig.~\ref{fig:AKopti}(f).


\AK{
\section{Discussion}
}



\AK{We discuss the stability of particles around the centerline $y=0$ in the presence of a background flow and confining walls if they are converging with a non-zero slope $\alpha$. This stability is in contrast to a channel with parallel walls, where an active particle swims away from the centerline provided that its orientation angle $\varphi$ is different from $n\pi$, $n=0,\pm1,\pm2,\ldots$.}

\begin{figure}[ht]
	\begin{center}
		\includegraphics[width=\textwidth]{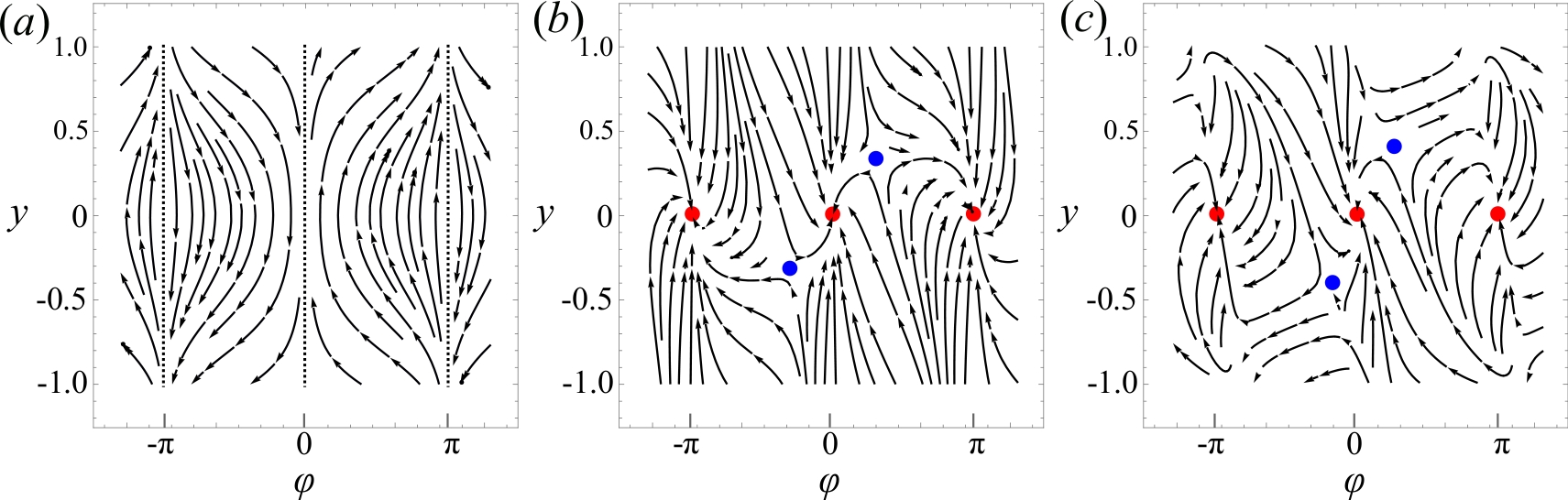}
		\caption{\footnotesize \AK{Phase portraits $(\varphi,y)$ for $v=0.2$, $H=1.0$ and $u_0=0.6$. (a)  System \eqref{varphi_poiseuille}-\eqref{y_poiseuille}, describing Poiseuille flow in a straight channel; dashed lines consist of stationary points. (b) System \eqref{varphi_convergent_simple}-\eqref{y_convergent_simple} describing a simplified convergent flow with $\alpha=0.25$,;  stationary points: stable $(\pi n,0)$ (in red) and pairs of  saddles with non-zero $y$ (in blue). Trajectories near the centerline converge to a stationary solution in the centerline. (c) System \eqref{orig-location}-\eqref{orig-orientation-angle} with the convergent flow ${\bf u}_{\text{BG}}=(u_x,u_y)$ used in Section~\ref{sec:focusing} with $x=-H/\alpha=-4.0$.}}
		\label{fig:stability}
	\end{center}	
\end{figure}

Indeed, in the case of a straight channel, $\alpha = 0$, the background flow is defined as $u_x=u_0 (H^2-y^2)$, $u_y=0$ (Poiseuille flow; $u_0$ is the strength of the flow, $2H$ is the distance between the walls). 
Then the system \eqref{orig-location}-\eqref{orig-orientation-angle} reduces to 
\begin{eqnarray}
	\dot{\varphi} &=& u_0 y (1-\cos 2\varphi) \label{varphi_poiseuille}\\
	\dot{y} &=& v_{0}\sin \varphi. \label{y_poiseuille}
\end{eqnarray} 
Here we omit the equation for $x(t)$ due to invariance of the infinite channel with respect to $x$ and neglect orientation fluctuations, that is $D_r=0$.
The phase portrait for this system is depicted in Fig.~\ref{fig:stability}(a). Dashed vertical lines $\varphi=n\pi$, $n=0,\pm1,\pm2,\ldots$ consist of stationary solutions: if an active particle is initially oriented parallel to the walls, it keeps swimming parallel to them. If initially $\varphi$ is different from $n\pi$, then the active particle swims away from the centerline, $y(t)\to\pm\infty$ as $t \to \infty$. 

When the walls are converging, $\alpha > 0$, the $y$-component of the background flow is non-zero and directed towards the centerline. For the sake of simplicity we take $u_y=-\alpha y$, $\alpha>0$ and $u_x$ as in the Poiseuille flow, $u_x=u_0 (H^2-y^2)$. In this case, the system \eqref{orig-location}-\eqref{orig-orientation-angle} reduces to 
\begin{eqnarray}
	\dot{\varphi} &=& -(\alpha/2)\sin 2\varphi + u_0 y (1-\cos 2\varphi) \label{varphi_convergent_simple}\\
	\dot{y} &=& -\alpha y+ v_{0}\sin \varphi. \label{y_convergent_simple}
\end{eqnarray} 
The corresponding phase portrait for this system is depicted in Fig.~\ref{fig:stability}(b). Orientations $\varphi=n\pi$ represent stationary solutions only if $y=0$. In contrast to the Poiseuille flow in a straight channel, see Eqs. (\ref{varphi_poiseuille}) and (\ref{y_poiseuille}),  these stationary solutions $(\varphi=\pi n, y=0)$ are asymptotically stable with a decay rate $\alpha$ (recall that $\alpha$ is the slope of walls). In addition to these stable stationary points there are pairs of unstable (saddle) points with non-zero $y$ (provided that $v_{0}>0$). In these saddle points, the distance from centerline $|y|$ does not change, since  a particle is oriented away from centerline, so the propulsion force moves the particle away from the centerline and this force is balanced by the convergent component of the background flow, $u_y$, moving the particle toward the centerline. The orientation angle $\varphi$ does not change since the torque from the Poiseuille component of the background flow, $u_x$, is balanced by the torque from the convergent component, $u_y$.       

We also draw the phase portrait for the converging flow ${\bf u}_{\text{BG}}=(u_x,u_y)$ introduced in Section~\ref{sec:model}, Fig.~\ref{fig:stability}(c). 
One can compare the phase portraits Fig.~\ref{fig:stability}(b) and Fig.~\ref{fig:stability}(c) around the stationary point $(\varphi=0, y=0)$ to see that the qualitative picture is the same: this stationary point is stable and it neighbors with two saddle points. 

{\AKn{The asymptotic stability of  $(\varphi=0, y=0)$ means that if a particle is close to the centerline and its orientation angle is close to $0$ (particle is oriented towards the outlet), it will keep swimming at the centerline pointing toward the outlet}, whereas in Poiseuille flow the particle would swim away. The asymptotic stability of $(\varphi=\pm \pi, y=0)$ is  evidence of that in the converging flow there is rheotaxis not only at walls but also in the bulk, specifically at the centerline.
Another consequence of this stability is the reduction of effective rotational diffusion of an active particle in the region around the centerline, that is the mean square angular displacement $\langle\Delta \varphi^2\rangle$ is bounded in time due to the presence of restoring force coming from the converging component of the background flow  (cf. diffusion quenching for Janus particles in \cite{DasGarCam2015}).}  
 \AK{Finally, we note that the nozzle has a finite length $L$ and thus, the conclusions of the stability analysis are valid if the stability relaxation time, $1/\alpha$ s, does not exceed the average dwell time $\langle T \rangle$. 
 We introduce a lower bound $\tilde{T}$ for the dwell time $\langle T\rangle$ as the dwell time of an active particle swimming along the centerline oriented forward, $\varphi=0$:} \AK{\begin{equation*}\tilde{T}=Lk/(\sigma v_0 (1-k))\ln |1+\sigma (1-k)/(k(\sigma+1))|.\end{equation*} }
 \AK{Our numerical simulations show that $\tilde{T}$ underestimates the average dwell time by a factor larger than two. Using this lower bound, we obtain the following sufficient condition for stability: $\dfrac{k w_{\text{in}}}{\sigma v_0}\ln\left|1+\dfrac{\sigma(1-k)}{k(\sigma+1)}\right|\geq 1$.  }    


\medskip 

\medskip 

\section{Conclusion}
In this work we study a dilute suspension of active rods in a viscous fluid extruded through a trapezoid nozzle. 
Using numerical simulations we examined the probability that a particle leaves the nozzle through the outlet - which is the result of the two counteracting phenomena. On the one hand, swimming downstream together with being focused by the converging flow increases the probability that an active rod leaves the nozzle at the outlet. On the other hand, rheotaxis results in a tendency of active rods to swim upstream. 

Theoretical approaches introduced in this paper can be used to design experimental setups for the extrusion of active suspensions through a nozzle. 
The optimal focusing is the result of a compromise. While for large flow rates it is very likely for active rods to leave the nozzle through the outlet very fast, their orientation is rather random and they pass through the outlet close to the walls. The particles are much better aligned with the flow for small flow rates and focused closer to the centerline of the nozzle, however the dwell time of the particles becomes quite large. Based on our findings the focusing is optimal if the velocity of the background flow and the self-propulsion velocity of the active rods are comparable. To reduce wall accumulation, the rods should have a small aspect ratio. 

\AK{We find that rheotaxis in bulk is possible for simple rigid rodlike active particles.} We also established analytically the local stability of active particle trajectories in the vicinity of the centerline. This stability leads to the decrease of the effective rotational diffusion of the active particles in this region \AK{as well as the emergence of rheotaxis away from walls.}
Our findings can be experimentally verified using biological or artificial swimmers in a converging flow.


\section*{Acknowledgements }

The work was supported by NSF DMREF grant DMS-1628411. A.K. gratefully acknowledges financial support through a Postdoctoral Research Fellowship (KA 4255/1-2) from the Deutsche Forschungsgemeinschaft (DFG). 

\section*{Author contributions statement}

Simulations have been performed by M.P. and A.K., the research has been conceived by L.B. and I.S.A. and all authors wrote the manuscript.

\bibliography{biblio}

\begin{thebibliography}{10}

\bibitem{Ram2010}
S.~Ramaswamy, ``The mechanics and statistics of active matter,'' {\em Annu.
  Rev. Condens. Matter Phys.}, vol.~1, no.~1, pp.~323--345, 2010.

\bibitem{MarJoaRamLivProRaoSim2013}
M.~Marchetti, J.~Joanny, S.~Ramaswamy, T.~Liverpool, J.~Prost, M.~Rao, and
  R.~Simha, ``Hydrodynamics of soft active matter,'' {\em Reviews of Modern
  Physics}, vol.~85, no.~3, p.~1143, 2013.

\bibitem{ElgWinGom2015}
J.~Elgeti, R.~Winkler, and G.~Gompper, ``Physics of microswimmers $-$ single
  particle motion and collective behavior: a review,'' {\em Reports on progress
  in physics}, vol.~78, no.~5, p.~056601, 2015.

\bibitem{DomCisChaGolKes2004}
C.~Dombrowski, L.~Cisneros, S.~Chatkaew, R.~Goldstein, and J.~Kessler,
  ``Self-concentration and large-scale coherence in bacterial dynamics,'' {\em
  Physical Review Letters}, vol.~93, no.~9, p.~098103, 2004.

\bibitem{sokolov2007concentration}
A.~Sokolov, I.~S. Aranson, J.~O. Kessler, and R.~E. Goldstein, ``Concentration
  dependence of the collective dynamics of swimming bacteria,'' {\em Phys. Rev.
  Lett.}, vol.~98, no.~15, p.~158102, 2007.

\bibitem{SurNedLeiKar2001}
T.~Surrey, F.~N\'ed\'elec, S.~Leibler, and E.~Karsenti, ``Physical properties
  determining self-organization of motors and microtubules,'' {\em Science},
  vol.~292, no.~5519, pp.~1167--1171, 2001.

\bibitem{SanCheDeCHeyDog2012}
T.~Sanchez, D.~Chen, S.~DeCamp, M.~Heymann, and Z.~Dogic, ``Spontaneous motion
  in hierarchically assembled active matter,'' {\em Nature}, vol.~491,
  no.~7424, pp.~431--434, 2012.

\bibitem{2005Riedel_Science}
I.~H. Riedel, K.~Kruse, and J.~Howard, ``A self-organized vortex array of
  hydrodynamically entrained sperm cells,'' {\em Science}, vol.~309,
  pp.~300--303, 2005.

\bibitem{Woolley}
D.~M. Woolley, ``Motility of spermatozoa at surfaces,'' {\em Reproduction},
  vol.~216, p.~259, 2003.

\bibitem{2008Friedrich_NJP}
B.~M. Friedrich and F.~J\"ulicher, ``The stochastic dance of circling sperm
  cells: sperm chemotaxis in the plane,'' {\em New J. Phys.}, vol.~10,
  p.~123035, 2008.

\bibitem{CavComGiaParSanSteVia2010}
A.~Cavagna, A.~Cimarelli, I.~Giardina, G.~Parisi, R.~Santagati, F.~Stefanini,
  and M.~Viale, ``Scale-free correlations in starling flocks,'' {\em
  Proceedings of the National Academy of Sciences}, vol.~107, no.~26,
  pp.~11865--11870, 2010.

\bibitem{CouKraJamRuxFra2002}
I.~Couzin, J.~Krause, R.~James, G.~Ruxton, and N.~Franks, ``Collective memory
  and spatial sorting in animal groups,'' {\em Journal of theoretical biology},
  vol.~218, no.~1, pp.~1--11, 2002.

\bibitem{VisZaf2012}
T.~Vicsek and A.~Zafeiris, ``Collective motion,'' {\em Physics Reports},
  vol.~517, no.~3, pp.~71--140, 2012.

\bibitem{BocquetPRL12}
I.~Theurkauff, C.~Cottin-Bizonne, J.~Palacci, C.~Ybert, and L.~Bocquet,
  ``Dynamic clustering in active colloidal suspensions with chemical
  signaling,'' {\em Phys. Rev. Lett.}, vol.~108, p.~268303, Jun 2012.

\bibitem{Bialke_PRL2013}
I.~Buttinoni, J.~Bialk\'e, F.~K\"ummel, H.~L\"owen, C.~Bechinger, and T.~Speck,
  ``Dynamical clustering and phase separation in suspensions of self-propelled
  colloidal particles,'' {\em Phys. Rev. Lett.}, vol.~110, p.~238301, Jun 2013.

\bibitem{Baskaran_PRL2013}
G.~S. Redner, M.~F. Hagan, and A.~Baskaran, ``Structure and dynamics of a
  phase-separating active colloidal fluid,'' {\em Phys. Rev. Lett.}, vol.~110,
  p.~055701, Jan 2013.

\bibitem{Palacci_science}
J.~Palacci, S.~Sacanna, A.~P. Steinberg, D.~J. Pine, and P.~M. Chaikin,
  ``Living crystals of light-activated colloidal surfers,'' {\em Science},
  vol.~339, p.~936, 2013.

\bibitem{wensink2012meso}
H.~Wensink, J.~Dunkel, S.~Heidenreich, K.~Drescher, R.~Goldstein, H.~L{\"o}wen,
  and J.~M. Yeomans, ``Meso-scale turbulence in living fluids,'' {\em
  Proceedings of the National Academy of Sciences}, vol.~109, no.~36,
  pp.~14308--14313, 2012.

\bibitem{SokAra2012}
A.~Sokolov and I.~Aranson, ``Physical properties of collective motion in
  suspensions of bacteria,'' {\em Physical Review Letters}, vol.~109,
  p.~248109, 2012.

\bibitem{SaiShe08}
D.~Saintillan and M.~Shelley, ``Instabilities and pattern formation in active
  particle suspensions: Kinetic theory and continuum simulations,'' {\em Phys.
  Rev. Lett.}, vol.~100, p.~178103, 2008.

\bibitem{RyaSokBerAra13}
S.~D. Ryan, A.~Sokolov, L.~Berlyand, and I.~Aranson, ``Correlation properties
  of collective motion in bacterial suspension,'' {\em New Journal of Physics,
  PMCID: PMC3878490}, vol.~15, p.~105021, 2013.
  \url{http://dx.doi.org/10.1088/1367-2630/15/10/105021}.

\bibitem{sokolov2009reduction}
A.~Sokolov and I.~S. Aranson, ``Reduction of viscosity in suspension of
  swimming bacteria,'' {\em Phys. Rev. Lett.}, vol.~103, no.~14, p.~148101,
  2009.

\bibitem{GacMinBerLinRouCle2013}
J.~Gachelin, G.~Mi\~no, H.~Berthet, A.~Lindner, A.~Rousselet, and E.~Cl\'ement,
  ``Non-{N}ewtonian {V}iscosity of {E}scherichia coli {S}uspensions,'' {\em
  Physical Review Letters}, vol.~110, p.~268103, 2013.

\bibitem{LopGacDouAurCle2015}
H.~L\'opez, J.~Gachelin, C.~Douarche, H.~Auradou, and E.~Cl\'ement, ``Turning
  bacteria suspensions into superfluids,'' {\em Physical Review Letters},
  vol.~115, no.~2, p.~028301, 2015.

\bibitem{HaiAraBerKar08}
B.~Haines, I.~Aranson, L.~Berlyand, and D.~Karpeev, ``Effective {V}iscosity of
  dilute bacterial suspensions: A two-dimensional model,'' {\em Physical
  Biology}, vol.~5, p.~046003 (9pp), 2008.

\bibitem{HaiSokAraBer09}
B.~M. Haines, A.~Sokolov, I.~S. Aranson, L.~Berlyand, and D.~A. Karpeev, ``A
  three-dimensional model for the effective viscosity of bacterial
  suspensions,'' {\em Phys. Rev. E}, vol.~80, p.~041922, 2009.

\bibitem{HaiAroBerKar10}
B.~M. Haines, I.~S. Aranson, L.~Berlyand, and D.~A. Karpeev, ``Effective
  {V}iscosity of bacterial suspensions: A three-dimensional pde model with
  stochastic torque,'' {\em Comm. Pure Appl. Anal.}, vol.~11, no.~1,
  pp.~19--46, 2010.

\bibitem{RyaHaiBerZieAra11}
S.~D. Ryan, B.~M. Haines, L.~Berlyand, F.~Ziebert, and I.~S. Aranson,
  ``Viscosity of bacterial suspensions: {H}ydrodynamic interactions and
  self-induced noise,'' {\em Rapid Communication to Phys. Rev. E}, vol.~83,
  p.~050904, 2011.

\bibitem{sokolov2010swimming}
A.~Sokolov, M.~M. Apodaca, B.~A. Grzybowski, and I.~S. Aranson, ``Swimming
  bacteria power microscopic gears,'' {\em Proceedings of the National Academy
  of Sciences}, vol.~107, no.~3, pp.~969--974, 2010.

\bibitem{di2010bacterial}
R.~Di~Leonardo, L.~Angelani, D.~Dell’Arciprete, G.~Ruocco, V.~Iebba,
  S.~Schippa, M.~Conte, F.~Mecarini, F.~De~Angelis, and E.~Di~Fabrizio,
  ``Bacterial ratchet motors,'' {\em Proceedings of the National Academy of
  Sciences}, vol.~107, no.~21, pp.~9541--9545, 2010.

\bibitem{kaiser2014transport}
A.~Kaiser, A.~Peshkov, A.~Sokolov, B.~ten Hagen, H.~L{\"o}wen, and I.~S.
  Aranson, ``Transport powered by bacterial turbulence,'' {\em Physical review
  letters}, vol.~112, no.~15, p.~158101, 2014.

\bibitem{WuLib2000}
X.-L. Wu and A.~Libchaber, ``Particle {D}iffusion in a
  {Q}uasi-{T}wo-{D}imensional {B}acterial {B}ath,'' {\em Physical Review
  Letters}, vol.~84, no.~13, pp.~3017--3020, 2000.

\bibitem{SokGolFelAra2009}
A.~Sokolov, R.~Goldstein, F.~Feldchtein, and I.~Aranson, ``Enhanced mixing and
  spatial instability in concentrated bacterial suspensions,'' {\em Physical
  Review E}, vol.~80, p.~031903, 2009.

\bibitem{pushkin2014stirring}
D.~O. Pushkin and J.~M. Yeomans, ``Stirring by swimmers in confined
  microenvironments,'' {\em Journal of Statistical Mechanics: Theory and
  Experiment}, vol.~2014, no.~4, p.~P04030, 2014.

\bibitem{DenissenkoPNAS}
P.~Denissenko, V.~Kantsler, D.~J. Smith, and J.~Kirkman-Brown, ``Human
  spermatozoa migration in microchannels reveals boundary-following
  navigation,'' {\em Proc. Natl. Acad. Sci. U.S.A.}, vol.~109, no.~21,
  pp.~8007--8010, 2012.

\bibitem{Chaikin2007}
P.~Galajda, J.~Keymer, P.~Chaikin, and R.~Austin, ``A wall of funnels
  concentrates swimming bacteria,'' {\em J. Bacteriol.}, vol.~189, p.~8704,
  2007.

\bibitem{ElgetiGompper13}
J.~Elgeti and G.~Gompper, ``Wall accumulation of self-propelled spheres,'' {\em
  Europhys. Lett.}, vol.~101, p.~48003, 2013.

\bibitem{Lee13Wall}
C.~F. Lee, ``Active particles under confinement: aggregation at the wall and
  gradient formation inside a channel,'' {\em New J. Phys.}, vol.~15, no.~5,
  p.~055007, 2013.

\bibitem{Ghosh}
P.~K. Ghosh, V.~R. Misko, F.~Marchesoni, and F.~Nori, ``Self-propelled {J}anus
  particles in a ratchet: Numerical simulations,'' {\em Phys. Rev. Lett.},
  vol.~110, p.~268301, Jun 2013.

\bibitem{Wensink2008}
H.~H. Wensink and H.~L\"owen, ``Aggregation of self-propelled colloidal rods
  near confining walls,'' {\em Phys. Rev. E}, vol.~78, p.~031409, Sep 2008.

\bibitem{BerTur1990}
H.~Berg and L.~Turner, ``Chemotaxis of bacteria in glass capillary arrays.
  {E}scherichia coli, motility, microchannel plate, and light scattering,''
  {\em Biophysical Journal}, vol.~58, no.~4, pp.~919--930, 1990.

\bibitem{RamTulPha1993}
M.~Ramia, D.~Tullock, and N.~Phan-Thien, ``The role of hydrodynamic interaction
  in the locomotion of microorganisms,'' {\em Biophysical Journal}, vol.~65,
  no.~2, pp.~755--778, 1993.

\bibitem{FryForBerCum1995}
P.~Frymier, R.~Ford, H.~Berg, and P.~Cummings, ``Three-dimensional tracking of
  motile bacteria near a solid planar surface,'' {\em Proceedings of the
  National Academy of Sciences: Biophysics}, vol.~92, pp.~6195--6199, 1995.

\bibitem{VigForWagTam2002}
M.~Vigeant, R.~Ford, M.~Wagner, and L.~Tamm, ``Reversible and {I}rreversible
  {A}dhesion of {M}otile {E}scherichia coli {C}ells {A}nalyzed by {T}otal
  {I}nternal {R}eflection {A}queous {F}luorescence {M}icroscopy,'' {\em Applied
  and Environmental Microbiology}, vol.~68, no.~6, pp.~2794--2801, 2002.

\bibitem{BerTurBerLau2008}
A.~Berke, L.~Turner, H.~Berg, and E.~Lauga, ``Hydrodynamic attraction of
  swinning microorganisms by surfaces,'' {\em Physical Review Letters},
  vol.~101, p.~038102, 2008.

\bibitem{Rot1963}
L.~Rothschild, ``Non-random distribution of bull spermatozoa in a drop of sperm
  suspension,'' {\em Nature}, vol.~198, no.~488, p.~1221, 1963.

\bibitem{YuaRaiBau2015}
J.~Yuan, D.~Raizen, and H.~Bau, ``A hydrodynamic mechanism for attraction of
  undulatory microswimmers to surfaces (bordertaxis),'' {\em Journal of the
  Royal Society Interface}, vol.~12, no.~109, p.~20150227, 2015.

\bibitem{DasGarCam2015}
S.~Das, A.~Garg, A.~Campbell, J.~Howse, A.~Sen, D.~Velegol, R.~Golestanian, and
  S.~Ebbens, ``Boundaries can steer active {J}anus spheres,'' {\em Nature
  Communications}, vol.~6, p.~8999, 2015.

\bibitem{JiaTorPeiBol2015}
Y.~Jiang, L.~Torrance, C.~Peichel, and D.~Bolnick, ``Differences in rheotactic
  responses contribute to divergent habitat use between parapatric lake and
  stream threespine stickleback,'' {\em Evolution}, vol.~69, no.~9,
  pp.~2517--2524, 2015.

\bibitem{HilKalMcMKos2007}
J.~Hill, O.~Kalkanci, J.~McMurry, and H.~Koser, ``Hydrodynamic surface
  interactions enable escherichia coli to seek efficient routes to swim
  upstream,'' {\em Physical review letters}, vol.~98, no.~6, p.~068101, 2007.

\bibitem{fu2012bacterial}
H.~C. Fu, T.~R. Powers, and R.~Stocker, ``Bacterial rheotaxis,'' {\em
  Proceedings of the National Academy of Sciences}, vol.~109, no.~13,
  pp.~4780--4785, 2012.

\bibitem{YuaRaiBau2015rheo}
J.~Yuan, D.~Raizen, and H.~Bau, ``Propensity of undulatory swimmers, such as
  worms, to go against the flow,'' {\em Proceedings of the National Academy of
  Sciences}, vol.~112, no.~12, pp.~3606--3611, 2015.

\bibitem{TouKirBerAra14}
M.~Tournus, A.~Kirshtein, L.~Berlyand, and I.~Aranson, ``Flexibility of
  bacterial flagella in external shear results in complex swimming
  trajectories,'' {\em Journal of the Royal Society Interface}, vol.~12,
  no.~102, p.~20140904, 2014.
\newblock \url{http://dx.doi.org/10.1088/1367-2630/15/3/035029}.

\bibitem{PalSacAbrBarHanGroPinCha2015}
J.~Palacci, S.~Sacanna, A.~Abramian, J.~Barral, K.~Hanson, A.~Grosberg,
  D.~Pine, and P.~Chaikin, ``Artificial rheotaxis,'' {\em Science Advances},
  vol.~1, no.~4, p.~e1400214, 2015.

\bibitem{Sano_PRL2010}
H.-R. Jiang, N.~Yoshinaga, and M.~Sano, ``Active motion of a {J}anus particle
  by self-thermophoresis in a defocused laser beam,'' {\em Phys. Rev. Lett.},
  vol.~105, p.~268302, Dec 2010.

\bibitem{paxton2004catalytic}
W.~F. Paxton, K.~C. Kistler, C.~C. Olmeda, A.~Sen, S.~K. St.~Angelo, Y.~Cao,
  T.~E. Mallouk, P.~E. Lammert, and V.~H. Crespi, ``Catalytic nanomotors:
  autonomous movement of striped nanorods,'' {\em Journal of the American
  Chemical Society}, vol.~126, no.~41, pp.~13424--13431, 2004.

\bibitem{HowsePRL2007}
J.~R. Howse, R.~A.~L. Jones, A.~J. Ryan, T.~Gough, R.~Vafabakhsh, and
  R.~Golestanian, ``Self-motile colloidal particles: From directed propulsion
  to random walk,'' {\em Phys. Rev. Lett.}, vol.~99, p.~048102, Jul 2007.

\bibitem{Bechinger_JPCM}
I.~Buttinoni, G.~Volpe, F.~K\"ummel, G.~Volpe, and C.~Bechinger, ``Active
  {B}rownian motion tunable by light,'' {\em J. Phys. Condens. Matter},
  vol.~24, p.~284129, 2012.

\bibitem{Baraban_SM2012}
L.~Baraban, M.~Tasinkevych, M.~N. Popescu, S.~Sanchez, S.~Dietrich, and O.~G.
  Schmidt, ``Transport of cargo by catalytic {J}anus micro-motors,'' {\em Soft
  Matter}, vol.~8, pp.~48--52, 2012.

\bibitem{bricard2013emergence}
A.~Bricard, J.-B. Caussin, N.~Desreumaux, O.~Dauchot, and D.~Bartolo,
  ``Emergence of macroscopic directed motion in populations of motile
  colloids,'' {\em Nature}, vol.~503, no.~7474, pp.~95--98, 2013.

\bibitem{bricard2015emergent}
A.~Bricard, J.-B. Caussin, D.~Das, C.~Savoie, V.~Chikkadi, K.~Shitara,
  O.~Chepizhko, F.~Peruani, D.~Saintillan, and D.~Bartolo, ``Emergent vortices
  in populations of colloidal rollers,'' {\em Nat. Commun.}, vol.~6, p.~7470,
  2015.

\bibitem{KaiserSciAdv2017}
A.~Kaiser, A.~Snezhko, and I.~S. Aranson, ``Flocking ferromagnetic colloids,''
  {\em Science Advances}, vol.~3, p.~e1601469, 2017.

\bibitem{EzhSai2015}
B.~Ezhilan and D.~Saintillan, ``Transport of a dilute active suspension in
  pressure-driven channel flow,'' {\em Journal of Fluid Mechanics}, vol.~777,
  pp.~480--522, 2015.

\bibitem{MalSta2017}
P.~Malgaretti and H.~Stark, ``Model microswimmers in channels with varying
  cross section,'' {\em The Journal of Chemical Physics}, vol.~146, no.~17,
  p.~174901, 2017.

\bibitem{Kirchhoff1996}
T.~Kirchhoff, H.~L\"owen, and R.~Klein, ``Dynamical correlations in suspensions
  of charged rodlike macromolecules,'' {\em Phys. Rev. E}, vol.~53,
  pp.~5011--5022, May 1996.

\bibitem{Jef1922}
G.~Jeffery, ``The {M}otion of {E}llipsoidal {P}articles {I}mmersed in a
  {V}iscous {F}luid,'' {\em Proceedings of the Royal Society A}, vol.~102,
  no.~715, pp.~161--179, 1922.

\bibitem{KimKar13}
S.~Kim and S.~J. Karrila, {\em Microhydrodynamics: principles and selected
  applications}.
\newblock Courier Dover Publications, 2013.

\bibitem{tirado}
M.~M. Tirado, C.~L. Martinez, and J.~G. de~la Torre, ``Comparison of theories
  for the translational and rotational diffusion-coefficients of rod-like
  macromolecules - application to short {DNA} fragments,'' {\em J. Chem.
  Phys.}, vol.~81, pp.~2047--2052, 1984.

\bibitem{zottl2013periodic}
A.~Z{\"o}ttl and H.~Stark, ``Periodic and quasiperiodic motion of an elongated
  microswimmer in poiseuille flow,'' {\em Eur. Phys. J. E}, vol.~36, no.~1,
  p.~4, 2013.

\bibitem{DreDunCisGanGol2011}
K.~Drescher, J.~Dunkel, L.~Cisneros, S.~Ganguly, and R.~Goldstein, ``Fluid
  dynamics and noise in bacterial cell-cell and cell-surface scattering,'' {\em
  Proceedings of the National Academy of Sciences}, vol.~108, no.~27,
  pp.~10940--10945, 2011.

\end{thebibliography}


\end{document}